\documentclass[twocolumn,showpacs,floats,floatfix,superscriptaddress,aps,pra]{revtex4-1}

\usepackage{amsfonts}
\usepackage{amssymb}
\usepackage{amsmath}
\usepackage{calc}
\usepackage{graphicx}
\usepackage{bm}

\usepackage[normalem]{ulem}
\usepackage{amsmath,amssymb}
\usepackage{multirow}
\usepackage{xcolor,soul}
\usepackage{srcltx}
\usepackage{hyperref,graphicx}

\def\be{ \begin{equation}}
\def\ee{ \end{equation}}
\def\bea{ \begin{eqnarray}}
\def\eea{ \end{eqnarray}}
\def\bse{ \begin{subequations}}
\def\ese{ \end{subequations}}
\def\bc{ \begin{center}}
\def\ec{ \end{center}}

\begin{document}

\author{Stefano Longhi$^{*}$} 
\affiliation{Dipartimento di Fisica, Politecnico di Milano, Piazza L. da Vinci 32, I-20133 Milano, Italy}
\affiliation{IFISC (UIB-CSIC), Instituto de Fisica Interdisciplinar y Sistemas Complejos, E-07122 Palma de Mallorca, Spain}
\email{stefano.longhi@polimi.it}

\title{Non-Hermitian skin effect and self-acceleration}
  \normalsize


%
\bigskip
\begin{abstract}
\noindent  
Non-Hermitian systems exhibit nontrivial band topology and a strong sensitivity of the energy spectrum on the boundary conditions. Remarkably,  a macroscopic number of bulk states get squeezed toward the lattice edges under open boundary conditions, an effect dubbed the non-Hermitian skin effect (NHSE). A well-established dynamical signature of the NHSE in real space is the  directional bulk flow (or persistent current) for arbitrary initial excitation of the system, which is observed at long times. Here we unravel a different dynamical signature of the NHSE in real space that manifests itself in the {\em early-time} dynamics of the system, namely self-acceleration of the wave function. Self-acceleration is demonstrated to occur rather generally in single--band lattice models probed by single-site excitation, where the acceleration turns out to be proportional to the area enclosed by the energy spectrum of the Bloch Hamiltonian under periodic boundary conditions. 
 The observation of wave packet self-acceleration at early times is a clear signature of the NHSE and should be experimentally accessible using synthetic non-Hermitian matter, for example in discrete-time photonic quantum walks.
 \end{abstract}

\maketitle


\section{Introduction}
The physical properties and topological phases of non-Hermitian (NH) periodic or quasi-periodic systems are attracting a great interest since the past few years \cite{r1,r1a,r2,r3,r4,r5,r6,r7,r8,r9,r10,r11,r12,r13,r14,r15,r16,r17,r18,r19,r20,r20a,r20b,r20c,r20d,r21,r22,r23,r24,r25,r26,r27,r28,r28a,r28b,r29,r30,r31,r32,r33,r34,r34a,r34b,r35,r36,r37,r38,r39,r40,r41,r42,r43,r44,r45,r46,r47,r48,r49,r49b,r50,r51,r51a,r51b,r51c,r51d,r51e,r51f,r51g,r51h,r51i,r51l,r52,r53,r54,r55,r56,r57,r58,r58a,r58b,r59,r60,r61,r62,r63,r64,r65,r66,r67,r67b,r67c,r68,r68a,r68b,r69,r70,r70b,Referee1,Referee2,r71,r71b,r71c,r72,r73,r74,r74a,r75,r76,r77,r78},  providing a newly emergent research area in condensed-matter physics and beyond (see e.g. the recent reviews \cite{r20,r1,r50,r51}). 
NH systems are non-conservative systems where the inherent loss and/or gain arising from interaction with an environment are described by a NH Hamiltonian. As an effective description, NH Hamiltonians find applications in different areas of physics \cite{r1}, ranging from optical, acoustical, mechanical and electrical systems to open quantum systems under continuous measurement and the description of quasiparticles in solids with disorder or interaction.\\
 A unique feature of certain NH lattices is the NH skin effect \cite{r5,r6,r7,r9,r29}, i.e. the strong dependence of the energy spectrum on the boundary conditions: while under  periodic boundary conditions (PBC) the energy spectrum describes one (or more) closed loop in complex plane enclosing a non-vanishing area $\mathcal{A}$, under open boundary conditions (OBC) the energy spectrum shrinks to one (or a set of) open arc in the interior of the PBC energy spectrum loop. Moreover, while under PBC the wave functions are extend Bloch modes like in ordinary Hermitian latices, under OBC an extensive number of bulk wave functions are exponentially localized at the edges, dubbed skin modes.  The origin of skin modes can be traced back to the nontrivial point-gap topology of the bulk energy spectra under PBC, thus establishing a bulk-edge correspondence for skin modes \cite{r3,r29}.  The NH skin effect greatly affects several dynamical properties of the system even in the bulk, including the appearance of a directional (chiral) bulk flow and persistent currents \cite{r22,r34,r68a,Referee1,Referee2,r78}, non-Bloch quench dynamics and dynamical quantum phase transitions \cite{r58a,r68b}, anharmonic Rabi oscillations \cite{r35}, chiral tunneling \cite{r32}, edge burst \cite{r70} and wave self-healing \cite{r71c}. In particular, the directional flow in the bulk of the lattice under rather arbitrary initial excitation conditions of the system, with an asymptotic (long-time) convective motion as dictated by a saddle-point analysis \cite{r22}, provides a feasible tool to detect the NH skin effect in bulk probing experiments \cite{Referee1,Referee2,r78}.\\
In this work we unveil a novel bulk dynamical signature of the NHSE, which arises in the {\em early} stage of time evolution: self-acceleration of the wave function. Such a phenomenon is demonstrated to arise rather generally in single-band lattice models for single-site initial excitation of the lattice. While the long-time dynamics of the wave function is dominated by the interference of  Bloch modes with the largest imaginary part of energy, leading to a directional flow of excitation along the lattice at a constant drift velocity in systems displaying the NHSE, the early-time dynamics of the wave function is dominated by the interference of the entire Bloch modes that constitute the wave function under single-site excitation, leading to a self (or transient) acceleration of the wave function in the early time evolution. The acceleration of the wave function is shown to be proportional to the area $\mathcal{A}$ enclosed by the PBC energy spectrum loop, thus vanishing in the absence of the NHSE. 
The predicted effect should be experimentally observable in synthetic classical or quantum NH lattices, for example in discrete-time photonic quantum walks \cite{r44,r48,Referee1,Referee2}.

\begin{figure}[t]
   \centering
    \includegraphics[width=0.49\textwidth]{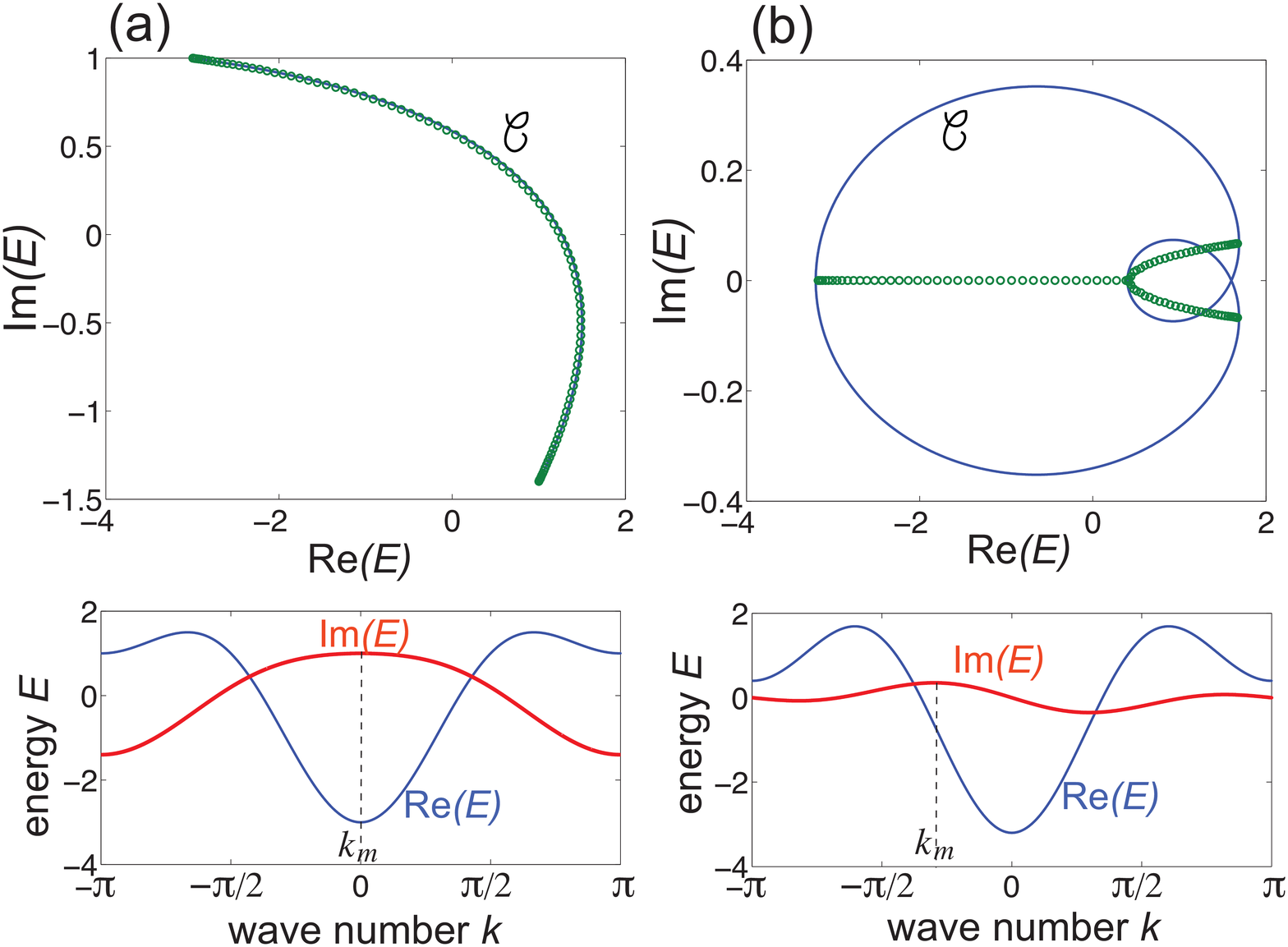}
    \caption{Energy spectra of a NH lattice with nearest and next-to-nearest neighbor hopping $t_{\pm1}$, $t_{\pm 2}$. In (a) $t_{-1}=t_1=1-0.6i$ and $t_{-2}=t_2=0.5+0.1i$, whereas in (b) $t_1=1$, $t_{-1}=0.8$, $t_2=0.8$, and $t_{-2}=0.6$. The upper panels show the PBC (solid curves) and OBC (open circles) energy spectra, while the lower panels depict the dispersion relation of the PBC energy curves $E=E_R(k)+iE_I(k)$.
    The NH lattice in (a) does not display the NHSE: the OBC and PBC energy spectra do coincide and are described by an open curve $\mathcal{C}$ in complex energy plane. The NH lattice in (b) displays the NHSE: the PBC energy spectrum describes a closed loop $\mathcal{C}$ with one self-intersection in complex plane, while the OBC energy spectrum is a set of open arcs in the interior of $\mathcal{C}$. The area $\mathcal{A}$ enclosed in the PBC energy loop is $\mathcal{A}= \pi (|t_{1}|^2+2 |t_2|^2-|t_{-1}|^2-2|t_{-2}|^2) \simeq -2.89$. The largest imaginary part of $E_I(k)={\rm Im} (E)$ is assumed at the Bloch wave number $k=k_m$.}
     \label{fig1}
\end{figure}

\section{Non-Hermitian lattice model and wave function dynamics}
\subsection{Hamiltonian, energy spectra and the non-Hermitian skin effect}
We consider a one-dimensional tight-binding lattice which is described, in the single band approximation, by the single-particle Hamiltonian in physical space
\begin{equation}
\hat{H}=\sum_{n,l} H_{n,l} |n \rangle \langle l |
\end{equation}
with matrix Hamiltonian $H_{n,l}=-t_{l-n}$, where $t_{\pm r}$ are the left/right hopping amplitudes between sites distant $r$ in the lattice. The Hamiltonian is not Hermitian whenever $t_{-r} \neq t_r^*$. The NHSE \cite{r6} arises rather generally in the presence of non-reciprocal hopping amplitudes, i.e. when $|t_{-r} | \neq |t_r|$, a prototypal example being the clean (disorder-free) Hatano-Nelson model \cite{r3,Hatano}. In this case the energy spectrum of $H$ is very sensitive to the boundary conditions. Indicating by 
\begin{equation}
P(\beta)=-\sum_l t_l \beta^l
\end{equation}
 the Laurent polynomial associated to the Toeplitz matrix $H$, for periodic boundary conditions the energy spectrum $\sigma(H_{PBC})$ reads $E_{PBC}(k)=P(\beta)$ with $\beta=\exp(ik)$ varying on the ordinary Brillouin zone $C_{\beta}$, i.e. $k$ real ($-\pi \leq k < \pi$) and $|\beta|=1$, while under open boundary conditions (OBC)  the energy spectrum $\sigma(H_{OBC})$ reads $E_{OBC}(k)=P(\beta)$ where now $\beta$ varies on the generalized Brillouin zone $\tilde{C}_{\beta}$, corresponding rather generally to $|\beta| \neq 1$, i.e. $k$ complex  \cite{r6,r9,r16,r29,r33,r69}. In the former case (PBC) the eigenfunctions of $H$ are the usual extended (Bloch) waves, while in the latter case (OBC)  they are exponentially localized toward either one of the two edges (skin modes), depending on whether $|\beta|>1$  or $|\beta| <1$.\\
  In the presence of the NHSE, $\sigma(H_{PBC})$ and $\sigma(H_{PBC})$ describe distinct curves in complex energy plane: while  $\sigma(H_{PBC})$ describes a closed loop $ \mathcal{C}$ with possible self-intersections,  $\sigma(H_{OBC})$ describes an open arc in the interior of the loop $\mathcal{C}$ \cite{r9,r16,r29}: only in the absence of the NHSE the $\sigma(H_{PBC})$ energy spectrum collapses to an open arc to coincide with $\sigma(H_{OBC})$ (with the exception of possible isolated points). Let us indicate by $\mathcal{A}$ the area enclosed by the loop $\mathcal{C}$, which is given by
\begin{eqnarray}
\mathcal{A} & = & \oint_{\mathcal{C}} E_I dE_R= \frac{1}{2} \oint_{\mathcal{C}} \left(E_I dE_R-E_R dE_I \right)  \nonumber \\
& = &    \frac{1}{2} \int_{-\pi}^{\pi} dk \left(E_I \frac{dE_R}{dk}-E_R \frac{dE_I}{dk} \right)   \label{area}
\end{eqnarray}
where $E_R(k)$ and $E_I(k)$ are the real and imaginary parts of the energy $E_{PBC}(k)$, respectively. Clearly, in a NH lattice without the NHSE the area $\mathcal{A}$ always vanishes, whereas in a NH lattice displaying the NHSE the area $\mathcal{A}$ is rather generally non-vanishing, with the very special exception of self-intersecting loops where the areas of distinct regions sum up with their own sign yielding an overall vanishing area. Therefore, rather generally a non-vanishing value of the area $\mathcal{A}$ can be regarded as a signature of the NHSE. Note that, since $E(k)=-\sum_n t_n \exp(ikn)$, in terms of hopping amplitudes $t_n$ of the lattice the area $\mathcal{A}$  takes the explicit form 
\begin{equation}
\mathcal{A}= - \pi  \sum_n n |t_n|^2
\end{equation}
which vanishes for reciprocal coupling ($|t_{-n}|=|t_n|$) but rather generally is non-vanishing for non-reciprocal (asymmetric) coupling. Interestingly, as shown in Ref.\cite{r34} a non-vanishing area $\mathcal{A}$ necessarily implies that, in the system under PBC, a non-vanishing persistent current $J=\oint_{\mathcal{C}} n(H,H^*)dH$ circulates in the ring. For specific forms of the distribution function $n(H,H^*)$, such a persistent current equals any area $\mathcal{A}' \leq A$ of any loop internal to $\mathcal{C}$ \cite{r34}.

As an example, Fig.1 shows the energy spectra under PBC and OBC in two NH lattice models with nearest and next-to-nearest hopping amplitudes $t_{\pm 1}$, $t_{\pm 2}$, one of which displays the NH skin effect [Fig.1(a)] while the other one does not [Fig.1(b)]. In the latter case the area $\mathcal{A}$ vanishes as $|t_{-1,-2}|=|t_{1,2}|$, while in the former case the energy loop $\mathcal{C}$ encloses rather generally a non-vanishing area $\mathcal{A}=- \pi (|t_1|^2-|t_{-1}|^2+2 |t_2|^2-2|t_{-2}|^2)$.

\subsection{Wave function dynamics}
We consider here the temporal dynamics of the wave function $|\psi(t) \rangle= \sum_n \psi_n(t) | n \rangle$ in an infinitely-extended lattice, i.e. far from any possible edge, with some given initial excitation condition $\psi_n(0)$ of the system at initial time $t=0$. The wave function satisfies the Schr\"odinger equation 
\begin{equation}
i \frac{\partial | \psi(t) \rangle}{\partial t} = \hat{H} |\psi(t) \rangle 
\end{equation}
on an infinitely-extended spatial domain, which can be readily solved in Fourier (momentum) space yielding
\begin{equation}
\psi_n(t)=\int_{-\pi}^{\pi} dk F(k) \exp[ikn-i E(k) t]. \label{soluzione}
\end{equation}
In the above equation, $E(k)=E_{PBC}(k)$ is the energy spectrum under PBC and 
\begin{equation}
F(k)=\frac{1}{2 \pi} \sum_{n=-\infty}^{\infty} \psi_n(0) \exp(-ikn) \label{spettro}
\end{equation}
is the spectral (Bloch) excitation amplitude, which is determined by the initial excitation condition of the system. Initial normalization of the wave function is assumed, i.e. $\sum_n | \psi_n(0)|^2=1$, corresponding to
\begin{equation}
 2 \pi \int_{-\pi}^{\pi} dk |F(k)|^2= 1.
\end{equation}

An interesting question of major relevance from both theoretical and experimental viewpoints is whether the temporal dynamics of the wave function can unravel the NHSE. Previous works focused on the long-time dynamics of the wave function, showing that  a wave function drift at some non-vanishing constant speed $v_m$ along the lattice, for rather arbitrary initial excitation condition of the system, is a clear signature of the NHSE \cite{r22,r68a,Referee2,r78}. Such a result has been successfully applied in bulk probing experiments of both classical and quantum NH systems to unveil the NHSE \cite{r68a,Referee1,Referee2,r78}. 
In the long-time limit ($t \rightarrow \infty$), from Eq.(\ref{soluzione}) it is clear that the dynamics is dominated by the interference of the extended Bloch modes with wave number $k$ close to $k_m$, at which the imaginary part of the PBC energy, $E_I(k)$, reaches its largest value. In fact, the growth rate of the wave function $\psi(t)=\psi_{n=vt}(t)$ along the space-time line $n=vt$, defined by the Lyapunov exponent
\begin{equation}
\lambda(v)=\lim_{t \rightarrow \infty} \log \frac{| \psi(t)|}{t},
\end{equation}
reaches its largest value, $\lambda=E_I(k_m)$, at the drift velocity $v=v_m=(dE_R/dk)_{k_m}$, which is non-vanishing in systems displaying the NHSE \cite{r22}. This result does not depend on the specific initial excitation condition of the system, and therefore its observation does not require any special initial preparation of the system.

\section{Wave function dynamics at early times: self-acceleration and the non-Hermitian skin effect}
\noindent
Here we focus our attention to the early time dynamics of the wave function, unveiling a novel bulk signature of the NHSE: self-acceleration. To this aim, let us indicate by $n_{CM}(t)$ the center of mass of the wave function at time $t$, which is given by the normalized mean value of the position operator
\begin{equation}
n_{CM}(t)= \frac{\langle \psi | n | \psi \rangle}{\langle \psi | \psi \rangle}=\frac{\sum_n n |\psi_n(t)|^2}{\sum_n | \psi_n(t)|^2}.
\end{equation}
Using Eq.(\ref{soluzione}), it can be readily shown that $n_{CM}(t)$ takes the form (see Appendix A for technical details)
\begin{equation}
n_{CM}(t)= \frac{ i \int_{-\pi}^{\pi} dk \left( F^* \frac{dF}{dk} -i |F(k)|^2 \frac{dE}{dk} t \right) \exp[2 E_I(k)t]}{\int_{-\pi}^{\pi} dk |F(k)|^2 \exp[2 E_I(k) t]  }  \label{CM}
\end{equation}
where $E(k)=E_R(k)+iE_I(k)$ is the dispersion curve of the PBC energy spectrum. According to the Lyapunov exponent analysis \cite{r22}, the asymptotic form of $n_{CM}(t)$ as $ t \rightarrow \infty$ is independent of the particular shape of $F(k)$, i.e. initial excitation of the system, and reads (see Appendix A)
\begin{equation}
n_{CM}(t) \sim v_m t \label{tlong}
\end{equation}
where $v_m=(dE_R/dk)_{k_m}$ is the drift velocity at the Bloch wave number $k=k_m$ corresponding to the largest value of $E_I(k)$, with $v_m \neq 0$ when the NH lattice displays the NHSE.\\
On the other hand, Eq.(\ref{CM}) indicates that the {\em short time} dynamics of $n_{CM} (t)$, i.e. as $t \rightarrow 0$, depends sensitively on the spectral amplitude $F(k)$, i.e. initial excitation of the system.
 To unravel the signature of the NHSE in the short-time dynamics of the wave function, we consider here a typical excitation condition of the lattice corresponding to single-site excitation, so that all extended Bloch states are equally excited.
In fact, for $\psi_n(0)= \delta_{n,0}$ the spectral amplitude $F(k)$ is flat [$F(k)=1/(2 \pi)$ according to Eq.(\ref{spettro})]. This kind of excitation enables to simply probe the existence of the NHSE in the lattice based on the early time dynamics of the system, as we are going to demonstrate. Assuming $F(k)=1/(2 \pi)$, the kinematic equation of the wave function center of mass [Eq.(\ref{CM})] reads
\begin{equation}
n_{CM}(t)=t  \frac{ \int_{-\pi}^{\pi} dk  \frac{dE}{dk} \exp[2 E_I(k)t]}{\int_{-\pi}^{\pi} dk  \exp[2 E_I(k) t]  } ,
\end{equation}
i.e.
\begin{equation}
n_{CM}(t)=t  \frac{ \int_{-\pi}^{\pi} dk  \frac{dE_R}{dk} \exp[2 E_I(k)t]}{\int_{-\pi}^{\pi} dk  \exp[2 E_I(k) t]  } \label{CMSS}
\end{equation}
since the integral 
\[
\int_{-\pi} ^{\pi} dk \frac{dE_I}{dk} \exp[2 E_I(k)t] =\frac{1}{2t} \int_{-\pi} ^{\pi}  dk \frac{d}{dk} \exp[2 E_I(k)t]
\]
vanishes. From Eq.(\ref{CMSS}) it readily follows that, in any Hermitian lattice or in an NH lattice that does not display the NH skin effect, one has $n_{CM}(t) \equiv 0$, i.e. the center of mass is at rest. In fact, in the Hermitian limit $E_I(k)=0$ and from Eq.(\ref{CMSS}) one obtains
\begin{equation}
n_{CM}(t)=t  \frac{ \int_{-\pi}^{\pi} dk  \frac{dE_R}{dk} }{\int_{-\pi}^{\pi} dk }=0
\end{equation}
consistent with the fact that in any Hermitian lattice the motion of the center of mass, in the absence of any gradient force, is uniform, and specifically at rest when the initial distribution of particle momentum has a vanishing mean value (like in the single-site lattice excitation). Likewise, in a NH lattice that does not display the NH skin effect one has $n_{CM}(t) \equiv 0$. To prove this result, let us observe that in this case  the loop $\mathcal{C}$ describing the PBC energy spectrum $E(k)$ degenerates to an open arc \cite{r9} and thus, for any Bloch wave number $k$ in the first Brillouin zone one can find another Bloch wave number $k^{\prime}$ in the same zone such that $E(k)=E(k^{\prime})$ and $dE(k)=-dE(k^{\prime})$. This implies that the integral
\[
\int_{-\pi}^{\pi} dk  \frac{dE_R}{dk} \exp[2 E_I(k)t]= \oint_{\mathcal{C}} dE_R \exp[2 E_I(k)t],
\]
\begin{figure}[t]
   \centering
    \includegraphics[width=0.49\textwidth]{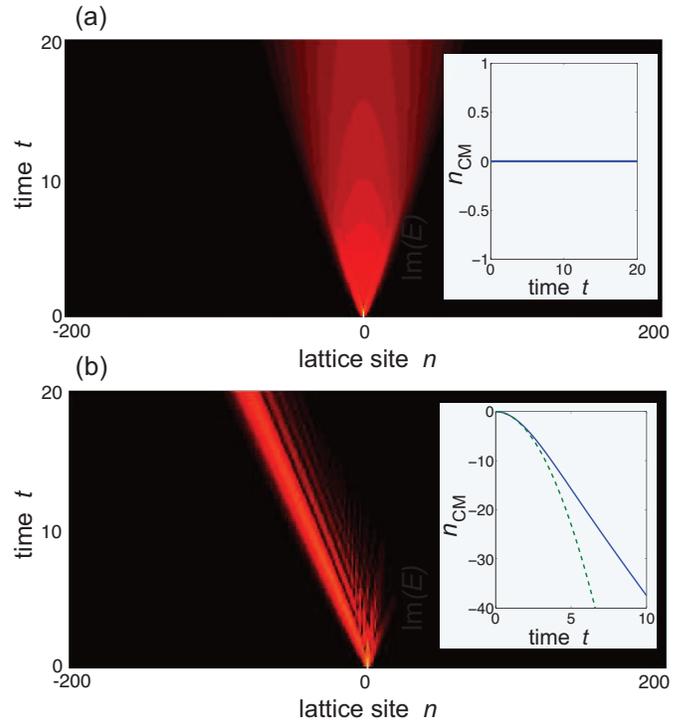}
    \caption{Time evolution of the wave function in the two NH lattices of Fig.1 for initial single-site excitation $\psi_n(0)=\delta_{n,0}$. The main panels show on a pseudo-color map the evolution of normalized wave function $|\tilde{\psi}_n(t)|=
   | \psi_n(t) |/  \sqrt{\sum_n | \psi_n(t)|^2}$, while the insets depict the corresponding time evolution of the wave function center of mass $n_{CM}(t)$. In (a) the NH lattice does not show the NHSE and the center of mass is at rest. In (b) the NH lattice displays the NHSE and the short time behavior of $n_{CM}(t)$ is parabolic with a constant acceleration according to Eq.(\ref{uffa}) of the main text. The dashed curve in the inset of (b) depicts the parabolic motion as predicted by Eq.(\ref{uffa}). }
     \label{fig2}
\end{figure}
and thus $n_{CM}(t)$, vanishes at any time $t$. Conversely, in a NH lattice displaying the NHSE the curve $\mathcal{C}$ describes a closed loop, with possible self-intersections, that does not degenerate to an open arc, so that $n_{CM}(t)$ is non-vanishing. It is interesting to calculate the short-time behavior of $n_{CM}(t)$, which is obtained from Eq.(\ref{CMSS}) in the $t \rightarrow 0$ limit. Introducing the Taylor expansion of the exponential term $\exp[2 E_I(k)t]=1+2 E_I(k)t+...$ in Eq.(\ref{CMSS}),  at leading order in $t$ one obtains
\begin{equation}
n_{CM}(t) \simeq \frac{1}{\pi} t^2 \int_{-\pi}^{\pi} dk  \frac{dE_R}{dk} E_I(k)= \frac{1}{\pi} t^2 \oint_{\mathcal{C}}  E_I dE_R.
\end{equation}
Taking into account that
\begin{equation}
\oint_{\mathcal{C}}  E_I dE_R =\frac{1}{2} \oint_{\mathcal{C}} (E_I dE_R-E_R dE_I)=\mathcal{A},
\end{equation}
where $\mathcal{A}$ is the area enclosed by the curve $\mathcal{C}$ [Eq.(\ref{area})], one finally obtains
\begin{equation}
n_{CM}(t) \simeq  \frac{1}{\pi} \mathcal{A}t^2. \label{uffa} 
\end{equation}
Equation (\ref{uffa}) shows that the early time dynamics of the wave function is characterized by a constant acceleration
\begin{equation}
a=\frac{2}{\pi} \mathcal{A}= 2  \sum_n n |t_n|^2 \label{acce}
\end{equation}
which is proportional to the area $\mathcal{A}$ enclosed by the PBC energy loop $\mathcal{C}$.
The acceleration is not due to any external gradient force, but to the asymmetric nature of hopping amplitudes and thus, ultimately, to the the NHSE. For such a reason, it can be referred to as a self-acceleration, or transient acceleration. Clearly, at longer times the acceleration decreases until to asymptotically vanish, corresponding to the long-time drift dynamical regime at the constant speed $v_m$. 
 The above results are illustrated in Fig.2, which depicts typical numerical results of the wave function time evolution in the two NH lattices of Fig.1 for initial single-site excitation of the system. The self-acceleration at early times is clearly visible in the lattice displaying the NHSE [Fig.2(b)], with an initial acceleration which is in excellent agreement with the theoretical prediction of Eq.(\ref{acce}).\par
 Two final comments are in order. 
 As a first comment, it should be mentioned that in single-band lattices self-acceleration of a wave packet can provide a clear signature of the NHSE even beyond the single-site excitation regime, provided that certain symmetry constraints are met. This point is discussed in the Appendix B. As a second comment, one could wonder whether the previous analysis can be extended to multi band systems. In this case the initial single-site (or more generally few-site) excitation of the lattice mixes the different bands so that the simple kinematical analysis is not anymore able to predict the NHSE from early-time self-acceleration arguments. However, for special initial excitation conditions, where a single band of the lattice is excited with the same amplitudes for all extended Bloch modes or even under special multi-band excitation, the previous analysis is still valid. An experimentally-relevant example of a two-band NH system, where the self-acceleration provides a signature of the NHSE, is discussed in the next section.

\section{Self-acceleration in non-Hermitian photonic quantum walks}
To illustrate the self-acceleration phenomenon, let us consider discrete-time NH photonic quantum walks as an experimentally-accessible platform of NH synthetic matter \cite{r44,r48,Referee1,Referee2,r78}. Specifically, we consider discrete-time quantum walks of optical pulses in coupled fiber loops that realize a synthetic mesh lattice \cite{fiber1,fiber2,fiber3}. The system consists of two fiber loops of slightly different lengths  $L \pm \Delta L$ (short and long paths) that are connected by a fiber coupler with a coupling angle $\beta$ [Fig.3(a)]. Balanced optical gain and loss $\pm h$ are applied in the short and long fiber loops, respectively. The traveling time of light in the two loops are $T \pm \Delta T$, where $T=L/c$,  $c$ is the group velocity of light in the fiber at the probing wavelength, and $\Delta T= \Delta L/c \ll T$ is the time mismatch arising from fiber length unbalance. After each round trip, the field amplitudes $u(t)$ and $v(t)$ of the light waves in the short and long loops at a given reference plane couple each other via the fiber coupler according to the time-delayed equations
\begin{eqnarray}
u(t+T-\Delta T) & = &  [\cos \beta u(t)+ i \sin \beta v(t) ]\exp(h) \label{F1} \\
v(t+T+\Delta T) & = &  [\cos \beta v(t)+ i \sin \beta u(t) ]\exp(-h). \label{F2}
\end{eqnarray}
We now consider the light dynamics at discretized times $t=t_n^m=n \Delta T+m T$, where $n=0, \pm 1, \pm2 ,...$ is the site number of the synthetic lattice at various time slots and $m$ is the round-trip number, assumed to match the traveling time $T$ along the mean path length $L$ [Fig.3(b)]. Note that the number $N$ of sites in in the lattice is limited by the constraint $N \Delta T <T$, i.e. $N<L/ \Delta L$.
Indicating by $u_n^{(m)}=u(t_n^m)$ and $v_n^{(m)}=v(t_{n+1}^m)$ the field amplitudes at the discretized times $t_n^m$, from Eqs.(\ref{F1}) and $(\ref{F2})$  it follows that
 light dynamics in the coupled fiber loops is governed by the discrete-time coupled equations 
 \begin{eqnarray}
 u^{(m+1)}_n & = & \left[   \cos \beta u^{(m)}_{n+1}+i \sin \beta v^{(m)}_{n}  \right]  \exp (h) \label{CME1} \\
 v^{(m+1)}_n & = & \left[   \cos \beta v^{(m)}_{n-1}+i \sin \beta u^{(m)}_{n}  \right] \exp (-h) \label{CME2}
 \end{eqnarray}
 which are the main equations describing the NH photonic quantum walk.
 \begin{figure}[t]
   \centering
    \includegraphics[width=0.45\textwidth]{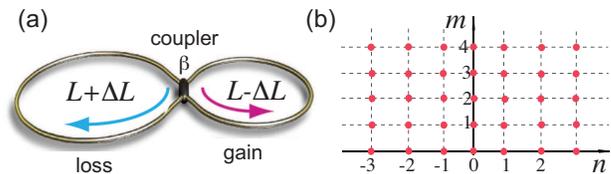}
    \caption{Discrete-time NH photonic quantum walk on a synthetic lattice. (a) Schematic of two coupled fiber loops with slightly length mismatch $L \pm \Delta L$. Optical gain is applied in the short fiber loop, while optical loss is applied to the long fiber loop. A fiber coupler with coupling angle $\beta$ mixes the light waves between the two fiber loops. (b) Schematic of the synthetic mesh lattice. The physical time $t$ is mapped at the discretized times  $t_n^m=n \Delta T+mT$, where $T=L/c$ is the mean travel time and $\Delta T= \Delta L /c \ll T$ is the travel time mismatch between the two fiber loops. The index $n$ corresponds to the site index in a synthetic one-dimensional spatial lattice, while the integer $m$ corresponds to a discrete time along which  the system evolves.}
     \label{fig3}
\end{figure}
  In a system under PBC, the eigenfunctions of Eqs.(\ref{CME1}) and (\ref{CME2}) are extended Bloch waves with wave number $k$ and quasi energy $\theta(k)$. Owing to the binary nature of the lattice, there are two quasi energy bands.  After letting
  \begin{equation}
  \left(
  \begin{array}{c}
  u_n^{(m)} \\
  v_n^{(m)}
  \end{array}
  \right)
  = \left(
  \begin{array}{c} 
  U \\
  V 
  \end{array}
  \right)
  \exp[ikn-i \theta(k)m] 
  \end{equation}
 one readily finds the following dispersion relations for the two quasi energy bands
 \begin{equation}
 \theta_{\pm}(k) = \pm {\rm {acos }} \left( \cos \beta \cos(k-ih) \right) \label{dispersion}
 \end{equation}
 with corresponding amplitudes of wave functions
 \begin{equation}
  \left(
  \begin{array}{c}
U \\
 V
  \end{array}
  \right)_{\pm}
  = \left(
  \begin{array}{c} 
  1 \\
  \frac{\exp(-h - i \theta_{\pm}(k))-\cos \beta \exp(ik)}{i \sin \beta} 
  \end{array}
  \right). \label{BF}
  \end{equation}
 The most general solution to Eqs.(\ref{CME1}) and (\ref{CME2}) reads
  \begin{equation}
  \left(
  \begin{array}{c}
  u_n^{(m)} \\
  v_n^{(m)}
  \end{array}
  \right)
  = \sum_{l= \pm }\int_{-\pi}^{\pi} dk F_l(k) \left(
  \begin{array}{c} 
  U \\
  V 
  \end{array}
  \right)_l
  \exp[ikn-i \theta_l(k)m] 
  \end{equation}
 where the spectral amplitudes $F_{\pm}(k)$ are determined from the initial excitation condition of the system. To establish a connection between the discrete-time photonic quantum walk and the continuous-time dynamics in a single-band NH lattice described by a continuous Schr\"odinger equation, let us consider a coupling angle $\beta$ close to $\pi/2$, so that $| \cos \beta | \ll 1$ and $\sin \beta \sim 1$. In this limit, the dispersion relations Eq.(\ref{dispersion}) of the two quasi energy bands read 
 \begin{equation}
 \theta_{\pm} (k) \simeq  \pm \frac{\pi}{2} \mp E(k),
 \end{equation}
  where we have set
 \begin{equation}
 E(k)=\cos \beta \cos (k-ih).
 \end{equation}
Since $|E(k)| \ll 1$, the amplitudes of corresponding Bloch wave functions [Eq.(\ref{BF})] take the simple form
 \begin{equation}
  \left(
  \begin{array}{c}
U \\
 V
  \end{array}
  \right)_{\pm}
  \simeq \left(
  \begin{array}{c} 
  1 \\
 \mp \exp(-h )
  \end{array}
  \right)
  \end{equation}
 and thus the most general solution to Eqs.(\ref{CME1}) and (\ref{CME2}) reads
 \begin{widetext}
 \begin{equation}
  \left(
  \begin{array}{c}
  u_n^{(m)} \\
  v_n^{(m)}
  \end{array}
  \right)
  = (-i)^m \int_{-\pi}^{\pi} dk F_+(k) \left(
  \begin{array}{c} 
  1 \\
  -\exp(-h)
  \end{array}
  \right)
  \exp[ikn+i E(k)m] + (i)^m \int_{-\pi}^{\pi} dk F_-(k) \left(
  \begin{array}{c} 
  1 \\
  \exp(-h)
  \end{array}
  \right)
  \exp[ikn-i E(k)m]. \label{basta}
  \end{equation}
\end{widetext}
 By a proper initial preparation of the system, we can excite a single quasi energy band with a flat spectral amplitude. For example, let us consider the initial condition
 \begin{equation}
 u_n^{(0)}= \delta_{n,0} \; , \; \;  v_n^{(0)}= \exp(-h) \delta_{n,0} \label{IC}
 \end{equation}
 corresponding to two-pulse excitation of the system, with one pulse injected into the short-path loop and the other one, with reduced amplitude $\exp(-h)$, in the long-path loop at the same time slot $n=0$. In this case
 one has $F_+(k)=0$ and $F_-(k)=1/(2 \pi)$, i.e. only the $l=-$ quasi energy band is excited with a flat spectral amplitude. Correspondingly, $v_n^{(m)}=\exp(-h) u_n^{(m)}$ and $u_n^{(m)}=(i)^m \psi_n^{(m)}$, where we have set
 \begin{equation}
 \psi_n^{(m)}=\frac{1}{2 \pi} \int_{-\pi}^{\pi} \exp[ikn-i E(k) m].
 \end{equation}
 \begin{figure}[t]
   \centering
    \includegraphics[width=0.49\textwidth]{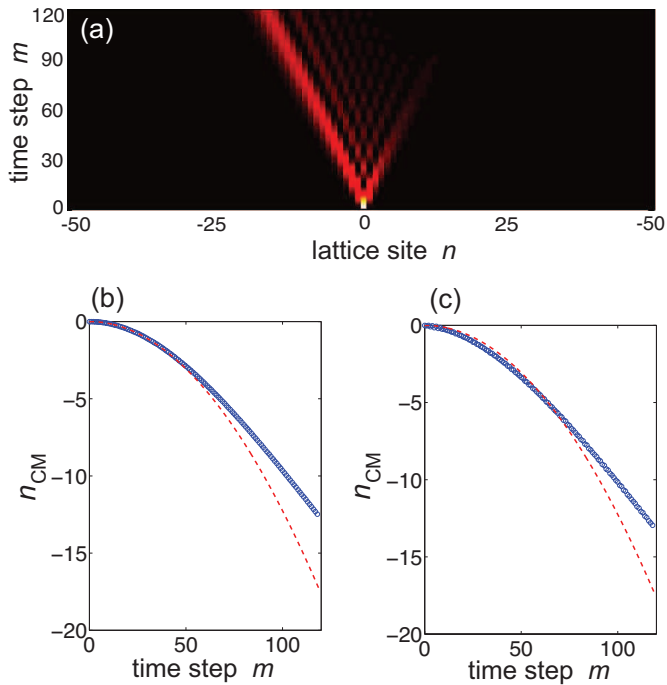}
    \caption{NH photonic quantum walk in a coupled fiber loop for a coupling angle $\beta= 0.9 \times \pi /2$ and a gain/loss parameter $h=0.05$. (a) Discrete-time evolution of the normalized light intensity $(|u_n^{(m)}|^2+|v_n^{(m)}|^2)/ \sum_n (|u_n^{(m)}|^2+|v_n^{(m)}|^2)$ at various lattice sites $n$ on a pseudo color map. Initial excitation condition of the system is $u_n^{(0)}=\delta _{n,0}$ and $v^{(0)}_n=\exp(-h) \delta_{n,0} \simeq \delta_{n,0}$. The corresponding evolution of the center of mass $n_{CM}(m)$ versus the discrete time step $m$ is shown in panel (b) by open circles. The dashed parabolic curve, almost overlapped with the open circles in the early stage of the dynamics, corresponds to the parabolic motion with an acceleration $a$ as predicted by the theoretical analysis [Eq.(\ref{thpredo})]. (c) Same as (b), but for the initial excitation condition $u_n^{(0)}=\delta _{n,0}$ and $v^{(0)}_n= 0$.}
     \label{fig2}
\end{figure}
 The center of mass of normalized light intensity in the mesh lattice at discrete time step $m$ reads
 \begin{equation}
 n_{CM}(m)=\frac{\sum_n n (|u_n^{(m)}|^2+|v_n^{(m)}|^2)}{\sum_n (|u_n^{(m)}|^2+|v_n^{(m)}|^2)}=\frac{\sum_n n |\psi_n^{(m)}|^2}{\sum_n |\psi_n^{(m)}|^2}. \label{CMfiber}
 \end{equation}
 Note that, since $| E(k)| \ll 1$, the amplitudes $\psi_n^{(m)}$ vary slowly with index $m$ and thus, in the continuum limit we may set $m=t$ and 
 \begin{equation}
 \psi_n(t)=\frac{1}{2 \pi} \int_{-\pi}^{\pi} \exp[ikn-i E(k) t].
 \end{equation}
  The previous equation, together with Eq.(\ref{CMfiber}), clearly shows that the center of mass of the light waves in the synthetic lattice exactly reproduces the dynamical behavior of a single-band NH lattice with Hamiltonian possessing the PBC energy spectrum $E(k)=\cos \beta  \cos(k-ih)$, which corresponds to the Hatano-Nelson model with asymmetric nearest-neighbor hopping amplitudes $t_{-1}=- (\cos \beta /2) \exp(-h)$ and $t_{1}=- ( \cos \beta /2) \exp(h)$. According to Eqs.(\ref{uffa}) and (\ref{acce}), the early time evolution of the wave function is parabolic with a self-acceleration given by 
  \begin{equation} 
  a= \frac{2}{\pi} \mathcal{A}=- \cos^2 \beta \sinh (2h). \label{thpredo}
  \end{equation}
 where $\mathcal{A}$ is the area enclosed by the ellipse describing the PBC energy spectrum of the Hatano-Nelson model. It should be mentioned that, for a small gain/loss parameter $h \ll 1$, the same parabolic motion can be approximately obtained even for initial single-pulse excitation of one fiber loop solely, i.e. for the initial condition $u_{n,0}=\delta_{n,0}$ and $v_{n,0}=0$, which can be experimentally more feasible than the above two-pulse excitation regime. In this case, even though both quasi energy bands of the lattice are excited,  the trajectory of the center of mass at early times remains approximately parabolic with an acceleration $a$ as given by Eq.(\ref{thpredo}) (see Appendix C for technical details).\\
  We checked the predictions of the theoretical analysis by direct numerical simulations of Eqs.(\ref{CME1}) and (\ref{CME2}) in physical space. The lattice is initially excited by two pulses, one injected in the short loop and the other one in the long loop, with relative amplitudes $\exp(-h)$, according to Eq.(\ref{IC}). Note that, for a small gain/loss parameter $h$, one can assume $\exp(-h) \simeq 1$, i.e. a symmetric excitation condition. Figures 4(a) and (b) show the numerically-computed discrete-time evolution of normalized light intensity $(|u_n^{(m)}|^2+|v_n^{(m)}|^2)/ \sum_n (|u_n^{(m)}|^2+|v_n^{(m)}|^2)$ at various lattice sites $n$, along with the behavior of the center of mass $n_{CM}(m)$, for a coupling angle $\beta=0.9 \times (\pi/2)$ and for a gain/loss parameter $h=0.05$.  The early time dynamics clearly shows self-acceleration with an acceleration $a$ in excellent agreement with the theoretical prediction given by Eq.(\ref{thpredo}) [Fig.4(b)]. For single-pulse excitation of the short-path fiber loop, a similar parabolic motion is also observed in the early stage of the dynamics [Fig.4(c)].

\section{Conclusions} 
The non-Hermitian skin effect, i.e. the strong sensitivity of the energy spectrum on the boundary conditions and the condensation of a macroscopic number of bulk states towards the lattice edges, is ubiquitous in non-Hermitian lattices with non-reciprocal couplings and rooted in the point-gap topology of the energy spectrum under periodic boundary conditions. 
Bulk dynamics can provide clear signatures of the NHSE, despite the absence of edge effects in the systems, and are thus of major interest both theoretically and experimentally. In this work we unveiled a clear signature of the NHSE in one-band NH lattices that is observed in the early time evolution of the system: self-acceleration. Under single-site excitation, the center of mass of the wave function in the lattice describes a parabolic path in the early stage of the dynamics with an acceleration that is proportional to the area enclosed by the PBC energy spectrum in complex plane. The early-time acceleration does not arise from any external force, but it is transiently induced by non-reciprocal hopping in the lattice and thus dubbed $^{\prime}$self-acceleration$^{\prime}$. We illustrated this phenomenon by considering in details NH photonic quantum walks, which realize synthetic NH matter in the temporal domain. 

The present study unravels a new and universal dynamical signature of the non-.Hermitian skin effect which, besides of major physical relevance, could  provide a useful tool to experimentalists. In fact, self-acceleration of a wave packet in the early time dynamics, as compared with directional flow in the long-time dynamics, can be easier to detect in an experiment, since it requires to probe the system only for a short time scale, thus avoiding detrimental effects such as decoherence in quantum matter or the need of controlling the system over long times.

\acknowledgments
The author acknowledges the Spanish State Research Agency, through the Severo Ochoa
and Maria de Maeztu Program for Centers and Units of Excellence in R\&D (Grant No. MDM-2017-0711).

\appendix
\begin{widetext}
\section{Wave function center of mass}
In this appendix we prove Eqs.(\ref{CM}) and (\ref{tlong}) given in the main manuscript. The wave function center of mass at time $t$ reads
\begin{equation}
n_{CM}(t)=\frac{\langle \psi (t) | n | \psi (t) \rangle}{\langle \psi (t) | \psi (t) \rangle}= \frac{\sum_n n |\psi_n(t)|^2}{\sum_n | \psi_n(t)|^2}
\end{equation}
where normalization of the wave function at any time $t$ is required owing  the non-conservative nature of the system.
Using the spectral representation Eq.(\ref{soluzione}) of the wave function, one has
\begin{equation}
\langle \psi(t) | \psi(t) \rangle=  \sum_n
\int dk dk^{\prime} F(k) F^*(k^{\prime}) \exp[i(k-k^{\prime})n] \exp[-i E(k) t+i E^*(k^{\prime})t ]. \label{A2}
\end{equation}
Taking into into account that $k, k^{\prime}$ vary in the range ($-\pi, \pi)$ and that
\begin{equation}
S(k-k^{\prime}) \equiv \sum_{n=-\infty}^{\infty} \exp[i(k-k^{\prime})n]= 2 \pi \delta(k-k^{\prime}) \label{A3}
\end{equation}
from Eq.(\ref{A2}) one obtains
\begin{equation}
\langle \psi(t) | \psi(t) \rangle=2 \pi \int_{-\pi}^{\pi} dk |F(k)|^2 \exp[2 E_I(k)t] \label{A4}
\end{equation}
where $E_I(k)$ is the imaginary part of the PBC energy $E(k)$. Likewise, one has
\begin{eqnarray}
\langle \psi(t) | n | \psi(t) \rangle & =  &  \sum_n
\int dk dk^{\prime} F(k) F^*(k^{\prime})  n \exp[i(k-k^{\prime})n] \exp[-i E(k) t+i E^*(k^{\prime})t ] \nonumber \\
& = & -i \int dk^{\prime} F^*(k^{\prime})  \exp[i E^*(k^{\prime})t ] \int dk F(k)  \frac{ \partial S(k-k^{\prime})}{\partial k} \exp[-i E(k) t ]
 \label{A5}
\end{eqnarray}
After integration by parts, from Eq.(\ref{A5}) and using Eq.(\ref{A3}) one obtains
\begin{equation}
\langle \psi(t) | n | \psi(t) \rangle =  2 \pi i \int dk^{\prime} F^*(k^{\prime})  \exp[i E^*(k^{\prime})t ] \frac{\partial}{\partial k^{\prime}} \left\{  F(k^{\prime})  \exp[-i E(k^{\prime})t ]  \right\} \label{A6}
\end{equation}
 Equation (A6) can be written in the equivalent form
 \begin{equation}
\langle \psi(t) | n | \psi(t) \rangle =  2 \pi i \int dk^{\prime} \left\{ F^*(k^{\prime})  \frac{d{F(k^{\prime})}}{d k^{\prime}} -i t |F(k^{\prime})|^2 \frac{d E(k^{\prime})}{d k^{\prime}}  \right\} \exp[2 E_I(k^{\prime})t]. \label{A7}
\end{equation}
From Eqs.(\ref{A4}) and (\ref{A7}) it then follows Eq.(\ref{CM}) given in the main text.\\
Let us finally calculate the asymptotic behavior of $\langle \psi(t) | \psi(t) \rangle$ and $\langle \psi(t) | n | \psi(t) \rangle$ at long times, i.e. as $t \rightarrow \infty$. Indicating by $k=k_m$ the Bloch wave number at which $E_I(k)$ reaches its largest value, i.e. 
\begin{equation}
(dE_I/dk)_{k_m}=0 \; , \; \; (d^2 E_I /dk^2)_{k_m} <0
\end{equation}
it is clear that in the $ t \rightarrow \infty$ limit the dominant contribution to the integrals on the right hand sides of Eqs.(\ref{A4}) and Eqs.(\ref{A7}) comes from the wave numbers $k$ close to $k_m$, and that the second term on the right hand sides of Eq.(\ref{A7}) becomes dominant over the first one owing to the secularly growing term $\sim t$. After letting $\xi=k-k_m$, $E(k) \simeq E(k_m)+(dE/dk)_{k_m} \xi +(1/2) (d^2 E /dk^2)_{k_m} \xi^2$ with $(dE/dk)_{k_m}=(dE_R/dk)_{k_m}$ real and $(d^2E_I/dk^2)_{k_m}<0$, one then obtains in the large $t$ limit
\begin{equation}
\langle \psi(t) | \psi(t) \rangle \sim 2 \pi |F(k_m)|^2 \exp[2 E_I(k_m)t]  \int_{-\infty}^{\infty} d \xi  \exp[ t \xi^2 (d^2 E_I/dk^2)_{k_m}] 
\end{equation}
 \begin{equation}
\langle \psi(t) | n | \psi(t) \rangle  \sim  2 \pi t   |F(k_m)|^2 \left(\frac{d E_R(k)}{d k} \right)_{k_m}  \exp[2 E_I(k_m)t] \int_{-\infty}^{\infty}  d \xi \exp[ t \xi^2 (d^2 E_I/dk^2)_{k_m}] 
\end{equation}
and thus
\begin{equation}
n_{CM}(t)=\frac{\langle \psi (t) | n | \psi (t) \rangle}{\langle \psi (t) | \psi (t) \rangle} \sim  v_m t
\end{equation}
in the large $t$ limit, where 
\begin{equation}
v_m=\left(\frac{d E_R(k)}{d k} \right)_{k_m}
\end{equation}
is the drift velocity.

\section{NH skin effect, self-acceleration and symmetry constraints}
In this Appendix we discuss under which general excitation conditions of the lattice an initial self-acceleration of the wave packet can provide a clear signature of the NHSE. In fact, rather generally it is possible to excite the system so as to observe an initial acceleration, despite the system does not display the NHSE.\\
For a rather arbitrary excitation of the system defined by the complex spectral amplitude $F(k)=H(k) \exp[i \varphi (k)]$ of modulus $H(k)$ and phase $\varphi(k)$, the instantaneous acceleration $a$ of the wave function in the early time dynamics, given by $a=\lim_{t \rightarrow  0^+}(d^2 n_{CM}/dt^2)$, can be readily calculated from Eq.(\ref{CM}) given in the main text and reads
\begin{eqnarray}
a & = & -8 \pi \int_{-\pi}^{\pi} dk H^2(k) \left( \frac{d \varphi} {dk} \right) E_I^2(k)+8 \pi \int_{-\pi}^{\pi} dk H^2(k) E_I(k) \left( \frac{dE_R}{dk} \right)+ 32 \pi^2 \left( \int_{-\pi}^{\pi}dk H^2(k) E_I(k) \right) \times \nonumber \\
& \times & \left[  \int_{-\pi}^{\pi} dk \left( \frac{d \varphi} {dk} \right)  H^2(k) E_I(k) + \int_{-\pi}^{\pi} dk E_R(k) H(k) \left( \frac{dH}{dk} \right) \right] \label{barba}
\end{eqnarray}
with the constraints $2 \pi \int_{-\pi}^{\pi} dk H^2(k)=1$ (normalization of the wave function at $t=0$) and $\int_{-\pi}^{\pi} dk (d \varphi / dk) H^2(k)=0$, corresponding to $n_{CM}(0)=0$. For single-site excitation of the system, corresponding to $H(k)=1/ ( 2\pi)$ and $\varphi(k)=0$, Eq.(\ref{barba}) reduces  to Eq.(\ref{acce}) given in the main manuscript, which relates the acceleration $a$ to the area $\mathcal{A}$ enclosed by the PBC energy spectrum in complex plane: thus a non-vanishing acceleration of the wave function at initial time is a clear signature of the NHSE. Here we wish to discuss under which more general conditions a non-vanishing acceleration $a$ at initial time indicates that the system displays the NHSE. To this aim, we write the PBC dispersion curve $E(k)=- \sum_l t_l \exp(ikl)$ in the equivalent form
\begin{equation}
E(k)=-t_0-\sum_{l \geq 1} \left[ (t_l+t_{-l}) \cos (kl)+i (t_l-t_{-l}) \sin (kl) \right]
\end{equation}
and consider two typical NH terms in the Hamiltonian, corresponding to different symmetries of the the energy spectrum $E(k)$. 
In the first case (case I) we assume $t_l$ real with $t_{-l} \neq t_l$ for some index $l$, corresponding to a NH lattice with non-reciprocal couplings leading to the NHSE. The symmetry of the energy spectrum is $E(-k)=E^*(k)$, i.e. $E_R(k)$ is even while $E_I(k)$ is odd under the inversion $k \rightarrow -k$. In the second case (case II) we assume $t_{l}$ complex and $t_{-l}=t_l$, corresponding to a NH lattice with complex (but reciprocal) hopping amplitudes, which does not display the NHSE.  The symmetry of the energy spectrum in this case reads $E(-k)=E(k)$, i.e. both $E_R(k)$ and $E_I(k)$ are even functions under the inversion $k \rightarrow -k$. Let us now consider an excitation of the system satisfying the minimal constraint $\varphi(k)=0$ and $H(-k)=H(k)$. From the odd/even symmetry of the functions under the sign of the integrals in Eq.(\ref{barba}), it readily follows that in case II, which does not display the NHSE, one always has $a=0$, whereas in case I, which does display the NHSE, one has
\begin{equation}
a  =  8 \pi \int_{-\pi}^{\pi} dk H^2(k) E_I(k) \left( \frac{dE_R}{dk} \right) \label{ancora?}
\end{equation}
which is rather generally non-vanishing. As an example, let us specialize the analysis to  the long-wavelength limit, where the spectral excitation amplitude $H(k)$ is narrow at around $k=0$ and $\psi_n(0)$ is a slowly-varying function of site position $n$. We can consider the Taylor expansions of $E_R(k)$ and $E_I(k)$ at around $k=0$, which in case I read $E_R(k)  \simeq E_R(0)+(1/2) (d^2 E_R/dk^2)_0k^2$ and $E_I(k) \simeq (dE_I/dk)_0k$. This basically corresponds to considering the long-wavelength (continuum) limit of the tight-binding Hamiltonian, i.e. $H(k) \simeq E_R(0)+i(dE_I/dk)_0k+(1/2) (d^2E_R/dk^2)_0 k^2$ in momentum space. In this case from Eq.(\ref{ancora?}) one obtains
\begin{equation}
a  \simeq 8 \pi  \left( \frac{dE_I}{dk}\right)_{k=0} \left( \frac{d^2E_R}{dk^2}\right)_{k=0}    \int_{-\pi}^{\pi} dk k^2 H^2(k) = - 8 \pi^2  \left( \sum_l l^2 t_l \right) \left( \sum_l l t_l \right) \int_{-\pi}^{\pi} dk k^2 H^2(k) 
\end{equation}
i.e. the acceleration is proportional to the variance of the momentum distribution of the wave function.

\section{Kinematic in photonic quantum walks under single-pulse excitation}
In the limit of a small gain/loss parameter $h \ll 1$, the discrete-time evolution of light field amplitudes $u_n^{(m)}$ and $v_n^{(m)}$ in the two fiber loops can be approximated as (see Eq.(\ref{basta}) of the main text)
 \begin{equation}
  \left(
  \begin{array}{c}
  u_n^{(m)} \\
  v_n^{(m)}
  \end{array}
  \right)
  \simeq (-i)^m \int_{-\pi}^{\pi} dk F_+(k) \left(
  \begin{array}{c} 
  1 \\
  -1
  \end{array}
  \right)
  \exp[ikn+i E(k)m] + (i)^m \int_{-\pi}^{\pi} dk F_-(k) \left(
  \begin{array}{c} 
  1 \\
  1
  \end{array}
  \right)
  \exp[ikn-i E(k)m]. \label{B1}
  \end{equation}
For single-pulse excitation of the system in the short fiber loop solely, i.e. for $u_n^{(0)}= \delta_{n,0}$ and $v_n^{(0)}=0$, the spectral amplitudes are given by $F_{+}(k)=F_-(k)=1/(4 \pi)$, indicating that both quasi energy bands are equally excited with spectrally-fat amplitudes. Hence, from Eq.(\ref{B1}) one obtains
\begin{equation}
u_n^{(m)}=\frac{(-i)^m}{2 \pi} 
\left\{
\begin{array} {cc}
\int_{-\pi}^{\pi} dk \exp(ikn) \cos [E(k)m] & m \; \; {\rm even} \\
i \int_{-\pi}^{\pi} dk \exp(ikn) \sin [E(k)m] & m \; \; {\rm odd} \\
\end{array}
\right. \label{B2}
\end{equation}
and
\begin{equation}
v_n^{(m)}=\frac{(-i)^{m+1}}{2 \pi} 
\left\{
\begin{array} {cc}
\int_{-\pi}^{\pi} dk \exp(ikn) \cos [E(k)m] & m \; \; {\rm odd} \\
i \int_{-\pi}^{\pi} dk \exp(ikn) \sin [E(k)m] & m \; \; {\rm even} \\
\end{array}
\right. \label{B3}
\end{equation}
where $E(k)=\cos \beta \cos(k-ih)$ is the energy band dispersion curve. From Eqs.(\ref{B2}) and (\ref{B3}), after straightforward calculations one obtains
\begin{eqnarray}
\sum_n \left( |u_n^{(m)}|^2+|v_n^{(m)}|^2   \right) & = & \frac{1}{2 \pi} \int_{-\pi}^{\pi} dk  \cosh [2E_I(k)m] \label{cazz1} \\
\sum_n n \left( |u_n^{(m)}|^2+|v_n^{(m)}|^2   \right) & = &  \frac{m}{2 \pi} \int_{-\pi}^{\pi} dk  \frac{dE_R}{dk}\sinh [2E_I(k)m] \label{cazz2}
\end{eqnarray}
To study the early time dynamics, let $\epsilon \equiv m | \cos \beta |$ and let consider discrete-time steps $m$ such that $\epsilon \ll 1$. In this case, we can expand the hyperbolic sine and cosine terms on the right hand sides of Eqs.(\ref{cazz1}) and (\ref{cazz2}) up to second order in $\epsilon$, i.e. $\cosh [2 mE_I(k)] \simeq 1+2 m^2E_I^2(k)$ and $\sinh [2 m E_I(k)] \simeq 2 m E_I(k)$. In this way, one obtains
\begin{eqnarray}
\sum_n \left( |u_n^{(m)}|^2+|v_n^{(m)}|^2   \right) & = & 1+O(\epsilon^2) \\
\sum_n n \left( |u_n^{(m)}|^2+|v_n^{(m)}|^2   \right) & = & \frac{m^2}{\pi} \int_{-\pi}^{\pi}dk E_I \frac{dE_R}{dk}  +o(\epsilon^2)=  \frac{m^2 \mathcal{A}}{\pi}+o(\epsilon^2)
\end{eqnarray}
and thus
\begin{equation}
n_{CM}(m)= \frac{\sum_n n \left( |u_n^{(m)}|^2+|v_n^{(m)}|^2   \right)}{\sum_n \left( |u_n^{(m)}|^2+|v_n^{(m)}|^2   \right)}=\frac{\mathcal{A}}{\pi} m^2+o(\epsilon^2).
\end{equation}
where $\mathcal{A}=\oint dE_R E_I$ is the area enclosed by the PBC energy loop $E(k)$ in complex plane [Eq.(\ref{area}) in the main text]. This proves that, in the early-time evolution of the system, the light center of mass in the mesh lattice describes a parabolic trajectory with an acceleration $a= 2 \mathcal{A} / \pi$.

\end{widetext}

\end{document}